\documentclass[fleqn,usenatbib]{mnras}

\usepackage{newtxtext,newtxmath}
\usepackage[T1]{fontenc}
\usepackage{ae,aecompl}

\usepackage{graphicx}	% Including figure files
\usepackage{amsmath}	% Advanced maths commands
\usepackage{amssymb}	% Extra maths symbols
\usepackage{amsfonts}
\usepackage{epsfig}
\usepackage{lipsum}
\usepackage{lmodern}
\usepackage{tabularx}
\usepackage{lineno}
\usepackage{caption}

\usepackage{booktabs}
\usepackage{eurosym}

\newcommand{\fermi}{\textit{Fermi}-LAT}
\newcommand{\ssim}{\mathord{\sim}}
\newcommand{\gray}{$\gamma$-ray}
\newcommand{\grays}{$\gamma$-rays}
\newcommand{\CenA}{Cen\,A}

\title[A search for CenA-like features in the spectra of \emph{Fermi}-LAT detected radio galaxies]{A search for Centaurus A-like features in the spectra of \emph{Fermi}-LAT detected radio galaxies}

\author[C. B. Rulten et al.]{
Cameron B. Rulten,$^{1}$\thanks{E-mail: cameron.b.rulten@durham.ac.uk}
Anthony M. Brown,$^{1}$
and Paula M. Chadwick,$^{1}$
\\
$^{1}$Centre for Advanced Instrumentation, Department of Physics, University of Durham, South Road, Durham, DH1 3LE, United Kingdom
}

\date{Accepted 2020 January 3. Received 2019 December 16; in original form 2019 September 6}

\pubyear{2020}

\begin{document}
\label{firstpage}
\pagerange{\pageref{firstpage}--\pageref{lastpage}}
\maketitle

% Abstract of the paper
%++++++++++++++++++++++++++++++++++++++++++++++++
%
%     ABSTRACT
%
%++++++++++++++++++++++++++++++++++++++++++++++++

\begin{abstract}

Motivated by the detection of a hardening in the \gray{} spectrum of the radio galaxy Centaurus\,A, we have analysed $\ssim10$ years of \fermi{} observations of 26 radio galaxies to search for similar spectral features. We find that the majority of the radio galaxies' \gray{} spectral energy distributions are best fitted with a simple power-law model, and no spectral hardening similar to that found in Centaurus\,A was detected. We show that, had there been any such spectral features present in our sample of radio galaxies, they would have been seen, but note that 7 of the radio galaxies (3C\,111, 3C\,120, 3C\,264, IC\,4516, NGC\,1218, NGC\,2892 and PKS\,0625-35) show evidence for flux variability on 6-month timescales, which makes the detection of any steady spectral features difficult. We find a strong positive correlation (r = 0.9) between the core radio power at 5\,GHz and the \gray{} luminosity and, using a simple extrapolation to TeV energies, we expect around half of the radio galaxies studied will be detectable with the forthcoming Cherenkov Telescope Array.
%%One object, 4C\,+39.12, is a good candidate for detection with current ground-based gamma-ray telescopes.%%

\end{abstract}

% Select between one and six entries from the list of approved keywords.
% Don't make up new ones.

%%%%%%%%%%%%%%%%%%%%%%%%%%%%%%%%%%%%%%%%%%%%%%%%%%

%%%%%%%%%%%%%%%%% BODY OF PAPER %%%%%%%%%%%%%%%%%%

%\linenumbers

%++++++++++++++++++++++++++++++++++++++++++++++++
%
%     SECTION 1: INTRODUCTION
%
%++++++++++++++++++++++++++++++++++++++++++++++++

\section{Introduction}
\label{sec:introduction}

In the recently released \fermi{} 4FGL 8--year point source catalog \citep{Fermi-4FGL-2019-ARXIV}, nearly $70\%$ of the objects are associated with a known astrophysical object. Of these associated sources, the majority are classified as blazars: either of type BL Lac ($\ssim22\%$) or blazar, classification unknown ($\ssim23\%$). Additionally, Flat Spectrum Radio Quasars (FSRQs) account for $\ssim13\%$ of the associated point sources in the 4FGL. Less than $1\%$ of all 4FGL sources are associated with radio galaxies. 

According to the unified model of active galactic nuclei (AGN) \citep{Urry:1995aa}, radio galaxies are radio--loud AGN that have jets beamed at large inclination angles with respect to the observer's line of sight, and are therefore sometimes termed misaligned blazars. Unlike blazars, therefore, the non-thermal radiation emitted by radio galaxies is only modestly beamed \citep{Rieger:2017}. Thus radio galaxies are very interesting targets because any detected \gray{} signal from these AGN is not dominated by the highly beamed jet emission, meaning it might be possible to disentangle jet and core emission or even to detect other potential sources of \gray{} emission.

The nearby, well-studied and (on \fermi{} timescales) non-variable object Centaurus\,A (\CenA) is an obvious target to search for emission which does not originate in the jet. A study of \cite{Brown-2017-PhRvD} revealed evidence for a new population of energetic particles near \CenA's core. This evidence was manifested as a statistically significant ($\mathrm{>5 \sigma}$) hardening in the Fermi-LAT \gray{} spectrum, with the spectral index changing from $\mathrm{\Gamma = 2.73~\pm~0.02}$ to $\mathrm{\Gamma = 2.29~\pm~0.07}$ at a break energy of $\mathrm{2.6~\pm~0.3}$ GeV.

The study of \CenA{} was motivated by a possible mechanism for the origin of the most energetic cosmic rays (CRs) in AGN. Such CRs extend to energies beyond $\mathrm{>10^{20} eV}$, making it difficult to explain the energy spectrum if CRs represent an accelerated thermal population. To solve this problem it might be possible to use dark matter (DM) annihilations as a source of non-thermal particles that can be further accelerated in astrophysical shocks. These shock--accelerated particles should produce a power-law spectrum for all energies, and the DM annihilation should produce a spectrum with a cut-off at the DM particle mass. The combination of these two effects should result in a characteristic spectrum \citep{Lacroix-2014-PhysRevD}.

The lack of variability in \CenA{}'s emission ruled out the possibility of jet-induced leptonic processes being responsible for the spectral feature and it was found that the $\gamma$-ray spectrum of \CenA{} was compatible with a very large localized enhancement (i.e. a spike) in the DM halo profile \citep{Brown-2017-PhRvD}. However, it was noted that a population of unresolved millisecond pulsars or another population of energetic particles could also be responsible for the emission above 2.6\,GeV. Recent results from the H.E.S.S. telescopes have resolved the emission above 100\,GeV, and suggest that the highest-energy emission from \CenA{} comes from a small, inner jet close to \CenA's core (\cite{Sanchez-TeVPA-2018}; \cite{HESS-CenA-2018AA}), which could well be the source of the population of energetic particles postulated in \cite{Brown-2017-PhRvD}.

The discovery of a spectral hardening of \CenA{}'s $\gamma$-ray spectrum provides the motivation to look at other radio galaxies detected with \fermi{} in order to search for similar spectral features. This work describes our analysis of a selection of such radio galaxies. In Section \ref{sec:observations} we highlight the \fermi{} observations used, our radio galaxy selection criteria, and the data analysis methods employed. In Section \ref{sec:results} we focus on the results of our \fermi{} analysis before discussing possible interpretations of our findings in Section \ref{sec:interpretation}.

%++++++++++++++++++++++++++++++++++++++++++++++++
%
%     SECTION 2: FERMI OBSERVATIONS & ANALYSIS
%
%++++++++++++++++++++++++++++++++++++++++++++++++

\section{\fermi{} observations and data analysis}
\label{sec:observations}
%%--------------------------------------------------
%% Data analysis methodology
%%--------------------------------------------------

The Large Area Telescope [LAT; \cite{Atwood:2009apj}] aboard the NASA \textit{Fermi} \gray{} Space Telescope is a wide--field pair conversion telescope sensitive to \grays{} over the approximate energy range $\mathrm{30~MeV \leqslant E \leqslant 300~GeV}$. The \fermi{}  was launched from the Kennedy Space Center on June 11, 2008 and started conducting science operations on 11th August 2008; it has thus recently celebrated 11 years of near uninterrupted service. The great majority of data taken by \fermi{} during this time has been in all-sky-survey mode. This observing mode scans the entire sky every $\ssim 180$ minutes and has produced the deepest extragalactic scan ever at \gray{} energies.

%++++++++++++++++++++++++++++++++++++++++++++++++
%
%     SUBSECTION 2.1: RADIO GALAXY SELECTION
%
%++++++++++++++++++++++++++++++++++++++++++++++++

\subsection{Radio galaxy selection}
\label{subsec:rg-selection} % used for referring to this section from elsewhere

%%--------------------------------------------------
%% Catalogs used for selecting radio galaxies
%%--------------------------------------------------
The 26 radio galaxies selected and listed in Table \ref{table:one} are those identified and categorized as radio galaxies in the \fermi{} 4FGL catalog \citep{Fermi-4FGL-2019-ARXIV}, excluding four well--studied, nearby radio galaxies: M\,87, the Perseus cluster galaxies NGC\,1275 and IC\,310, all of which have been found to exhibit significant flux variability at $\gamma$-ray energies (\citep{AitBenkhali-2019-AA,Brown-2011-MNRAS,Aleksic-2014-Sci} respectively) and Cen\,A \citep{Brown-2017-PhRvD}. We have included PKS\,0625-35 in the list of selected radio galaxies; however, as discussed in a recent paper \citep{Abdalla-2018-MNRAS}, there is evidence to suggest that this galaxy could be a BL Lac object and thus its classification as a ``misaligned blazar'' may need to be reconsidered. The majority of the radio galaxies in our study are classified as having a Fanaroff-Riley type I (FR\,I) morphology \citep{Fanaroff:1974aa}; there are 6 Fanaroff-Riley type 2 (FR\,II) galaxies, and one compact radio galaxy (FR\,0). 

Our selection does not, of course, represent a complete list of radio galaxies detected above $\mathrm{20\ MeV}$, as there is always the possibility that some of the $1500+$ unassociated sources in the 4FGL catalog may be radio galaxies.

\begin{table*}
\footnotesize
\begin{tabular}{| l | l | r | r | r | l | r | r |}
\hline
\fermi{} name & Assoc. name & \textit{l} (deg.) & \textit{b} (deg.) & z & Morphology & Variability index & $\sigma$ ($\sqrt{TS}$) \\ \hline
4FGL J0322.6-3712e & Fornax\,A & 240.16 & -56.68 & 0.0059 & FR I & 36.5 & 16.99 \\ \hline
4FGL J0057.7+3023 & NGC\,315 & 124.56 & -32.49 & 0.0164 & FR II & 21.0 & 9.03 \\ \hline
4FGL J0708.9+4839 & NGC\,2329 & 168.57 & 22.79 & 0.0197 & - & 4.0 & 7.23 \\ \hline
4FGL J0334.3+3920 & 4C\,+39.12 & 154.16 & -13.43 & 0.0203 & FR 0 & 20.1 & 8.81 \\ \hline
4FGL J1144.9+1937 & $\mathrm{3C\,264\dagger}$ & 235.72 & +73.03 & 0.0216 & FR I & 66.5 & 11.38 \\ \hline
4FGL J0931.9+6737 & NGC\,2892 & 145.14 & +39.87 & 0.0226 & - & 45.9 & 16.76 \\ \hline
4FGL J1630.6+8234 & NGC\,6251 & 115.76 & +31.19 & 0.0239 & FR I & 39.7 & 38.33 \\ \hline
4FGL J0009.7-3217 & IC\,1531 & 2.39 & -32.27 & 0.0256 & - & 38.1 & 7.46 \\ \hline
4FGL J2156.0-6942 & PKS\,2153-69 & 321.31 & -40.6 & 0.0280 & - & 22.2 & 8.80 \\ \hline
4FGL J0308.4+0407 & NGC\,1218 & 174.85 & -44.51 & 0.0288 & FR I & 49.1 & 20.15 \\ \hline
4FGL J1449.5+2746 & B2\,1447+27 & 41.25 & 63.87 & 0.0308 & - & 7.4 & 5.30 \\ \hline
4FGL J0433.0+0522 & 3C\,120 & 190.37 & -27.39 & 0.0336 & FR I & 306.8 & 24.56 \\ \hline
4FGL J0519.6-4544 & Pictor\,A & 251.59 & -34.63 & 0.0340 & FR II & 10.7 & 10.59 \\ \hline
4FGL J0758.7+3746 & NGC\,2484 & 182.67 & +28.82 & 0.0408 & FR I & 10.3 & 4.80 \\ \hline
4FGL J1454.1+1622 & IC\,4516 & 223.59 & +16.35 & 0.0452 & FR II & 73.3 & 13.71 \\ \hline
4FGL J0418.2+3807 & 3C\,111 & 161.67 & -08.81 & 0.0485 & FR II & 89.9 & 19.07 \\ \hline
4FGL J2341.8-2917 & PKS\,2338-295 & 355.36 & -29.31 & 0.0523 & - & 38.1 & 6.63 \\ \hline
4FGL J1516.5+0015 & PKS\,1514+00 & 1.38 & 45.98 & 0.0526 & FR II & 17.9 & 8.73 \\ \hline
4FGL J0627.0-3529 & $\mathrm{\operatorname{PKS\,0625-35}\dagger}$ & 243.45 & -19.96 & 0.0562 & FR I & 42.7 & 33.65 \\ \hline
4FGL J1306.3+1113 & TXS\,1303+114 & 316.05 & 73.71 & 0.0857 & FR I & 7.6 & 4.95 \\ \hline
4FGL J1518.6+0614 & TXS\,1516+064 & 8.86 & 49.25 & 0.1021 & FR I & 6.3 & 6.49 \\ \hline
4FGL J1843.4-4835 & PKS\,1839-48 & 347.17 & -18.72 & 0.1112 & FR I & 15.0 & 6.15\\ \hline
4FGL J1306.7-2148 & PKS\,1304-215 & 307.62 & 40.92 & 0.1260 & - & 22.8 & 11.85 \\ \hline
4FGL J2302.8-1841 & PKS\,2300-18 & 45.89 & -63.71 & 0.1289 & - & 18.0 & 9.44 \\ \hline
4FGL J1443.1+5201 & 3C\,303 & 90.52 & +57.50 & 0.1412 & FR II & 22.2 & 7.38 \\ \hline
4FGL J2326.9-0201 & PKS\,2324-02 & 351.72 & -02.03 & 0.1880 & - & 35.8 & 6.32 \\ \hline
\end{tabular}
\caption{Details of the radio galaxies analyzed in this study including their \fermi{} variability index and detection significance (obtained in this work). The radio galaxies were selected using the \fermi{} 4FGL catalog and are ordered by increasing redshift (z). A variability index $\mathrm{> 39.7}$ indicates a $\mathrm{< 1 \%}$ chance of being a steady source. The two TeV-detected radio galaxies are highlighted with a $\mathrm{\dagger}$.}
\label{table:one}
\end{table*}

%++++++++++++++++++++++++++++++++++++++++++++++++
%
%     SUBSECTION 2.2: DATA ANALYSIS
%
%++++++++++++++++++++++++++++++++++++++++++++++++

\subsection{Data analysis}
\label{subsec:data-analysis}

%%--------------------------------------------------
%% Results of the data analysis
%%--------------------------------------------------

Roughly 10 years worth of \fermi{} data were used for each target. The exact exposure spans MJD 54682.65527778 through to and including MJD 58362.0, which is equivalent to 317895388 \fermi{} seconds, 3679.34 \fermi{} days or 10.08 \fermi{} years. We consider photons with energies between $\mathrm{0.1 - 300\,GeV}$ within a $\mathrm{15^{\circ}}$ circular region of interest (ROI) centred on each radio galaxy target. These photons were obtained from \fermi{} sky-survey observations in accordance with the \textsc{pass8} data analysis criteria. The \fermi{} recommended quality cuts were used, including a zenith angle cut of $\mathrm{90^{\circ}}$ (to reduce \gray{} contamination originating from the Earth's limb), \texttt{(DATA\_QUAL>0)\&\&(LAT\_CONFIG==1)} and \texttt{abs}(\texttt{rock\_angle})$\mathrm{\,<52}$.

We used the open--source \texttt{Python} package \texttt{Fermipy} \citep{Wood-Fermipy-2017-ICRC} to facilitate analysis of \fermi{} data with the \fermi{} \texttt{Fermitools} (v1.0.1). The analysis used the \texttt{P8R3$\_$SOURCE$\_$V2} instrument response function and adopted the binned maximum-likelihood method \citep{Mattox:1996apj}. To estimate the background, we included sources within the region of interest (ROI) listed in the 4FGL catalog \citep{Fermi-4FGL-2019-ARXIV} along with the recommended Galactic (\texttt{gll$\_$iem$\_$v07.fits}) and isotropic diffuse (\texttt{iso$\_$P8R3$\_$SOURCE$\_$V2$\_$v1.txt}) templates provided with the \texttt{Fermitools}.

Since our study considered 10 years of \fermi{} observations, we must search for additional point sources of \grays{} not accounted for by the 8-year integrated catalogue of the 4FGL. To do this, we used the \texttt{find\_sources} algorithm in \texttt{Fermipy} to construct a significance map centred on each radio galaxy candidate\footnote{The significance map was constructed assuming a point source with an $E^{-2}$ spectrum.}. This TS map was used to identify additional point sources of $\gamma$-rays, with $\mathrm{TS\geqslant25}$, that were not accounted for in our initial model. These new point sources were modelled using a power-law fixed at the location ($\alpha_{J2000}$, $\beta_{J2000}$) of the peak excess, and a final likelihood fit was performed with the normalisation and spectral index of the new point sources free to vary. 

%++++++++++++++++++++++++++++++++++++++++++++++++
%
%     SECTION 3: RESULTS
%
%++++++++++++++++++++++++++++++++++++++++++++++++

\section{results}
\label{sec:results}
\begin{figure*}
\centering
\captionbox[Text]{Spectral energy distributions obtained for the radio galaxies: 3C\,311, 3C\,120, 3C\,264, 3C\,303, 4C\,+39.12, B2\,1447+27, Fornax\,A and IC\,1531. Apart from 3C\,120, all the radio galaxy SEDs in this subset are best-fitted with a simple power-law model. The binning scheme is 2 bins per decade and a $\mathrm{95\%}$ confidence--level upper limit is shown for bins where $\mathrm{\sqrt{TS} < 2 \sigma}$ and the number of \gray{} photons above background in each bin is $\mathrm{\gamma < 2}$. \label{fig:seds-setA}}{%
\includegraphics[width=0.9\textwidth]{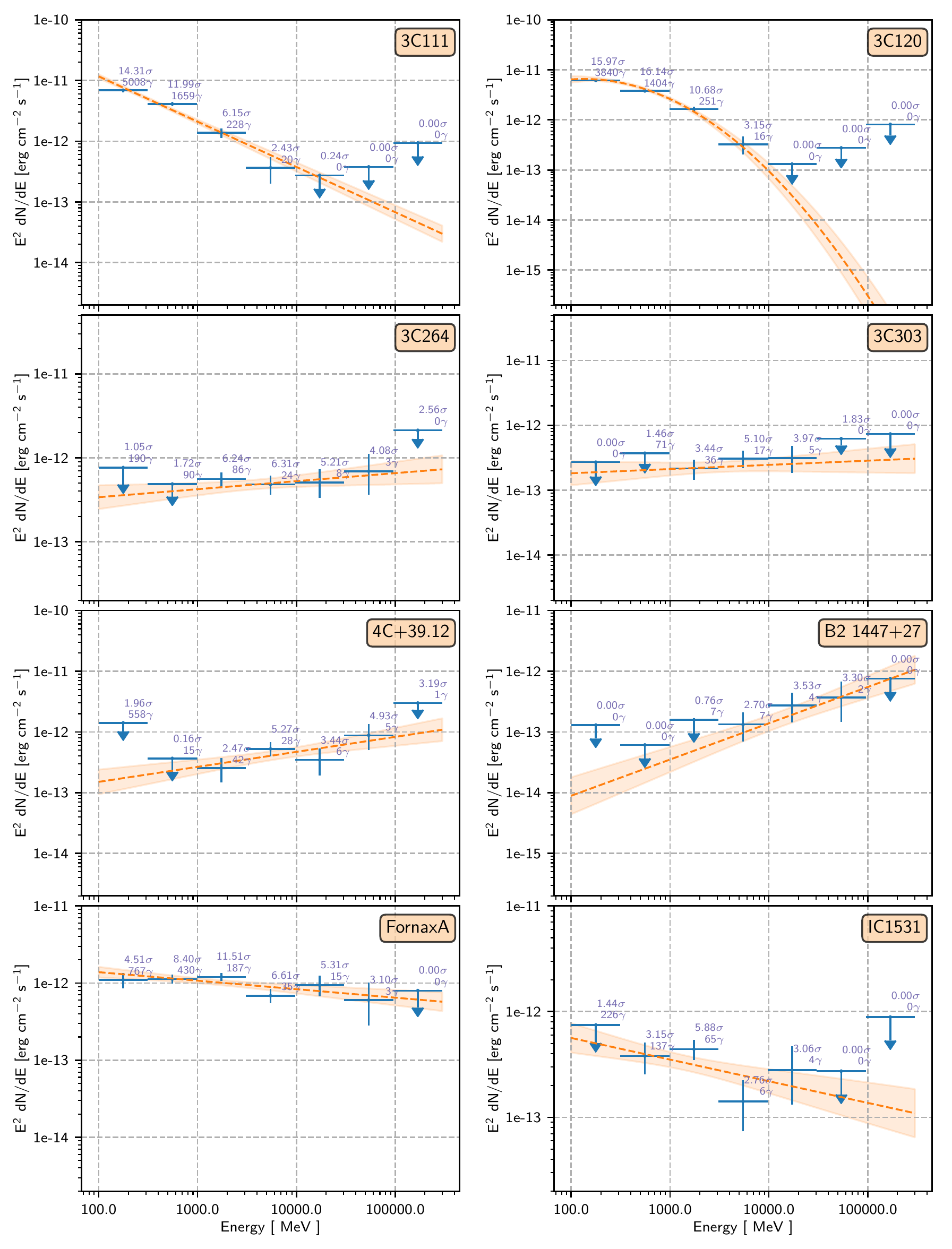}
}
\end{figure*}

\begin{figure*}
\centering
\captionbox[Text]{Spectral energy distributions obtained for the radio galaxies IC\,4516, NGC\,1218, NGC\,2329, NGC\,2484, NGC\,2892, NGC\,315, NGC\,6251 and PKS\,0625-35. Apart from NGC\,1218 and NGC\,6251, all the radio galaxy SEDs in this subset are best-fitted with a simple power-law model. The binning scheme is 2 bins per decade and a $\mathrm{95\%}$ confidence--level upper limit is shown for bins where $\mathrm{\sqrt{TS} < 2 \sigma}$ and the number of \gray{} photons above background in each bin is $\mathrm{\gamma < 2}$. \label{fig:seds-setB}}{%
\includegraphics[width=0.9\textwidth]{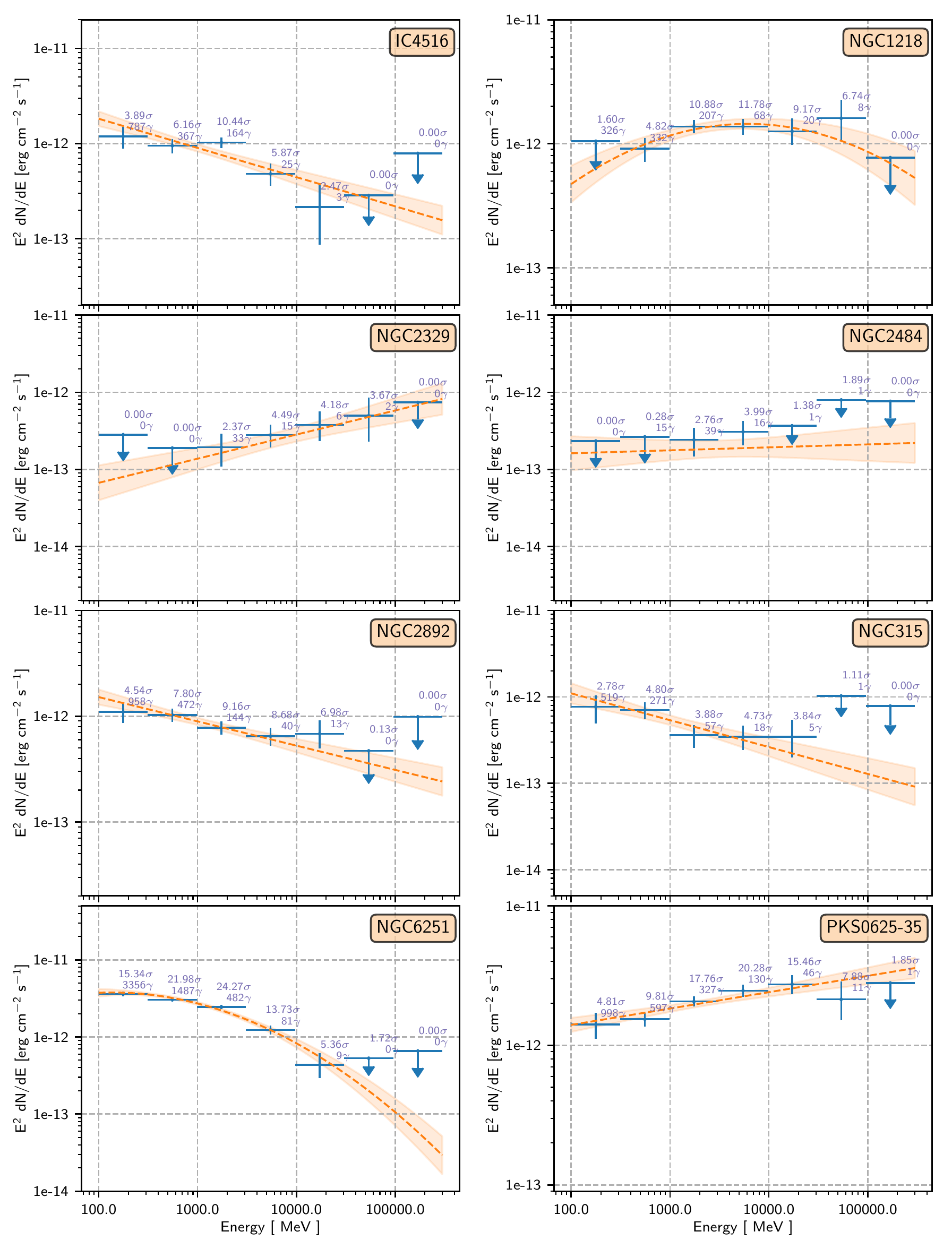}
}
\end{figure*}

Once all sources of $\gamma$-rays were accounted for in our data, we conducted temporal and spectral studies for all 26 radio galaxies considered in our research. For each, we produced a spectral energy distribution (SED); these are shown in Figures \ref{fig:seds-setA} and \ref{fig:seds-setB}. The SED flux points are generated using a separate likelihood analysis for each equally--spaced logarithmic energy bin. For each target we initially used 8 bins per decade, but then rebinned the flux data into a binning scheme of 2 bins per decade. Each spectral bin requires a statistical significance above background as defined by the test statistic, TS \footnote{Defined as twice the difference between the log-likelihoods of two different models, $2[logL- logL_0]$, where $L$ and $L_0$ are defined as the likelihoods of individual model fits \citep{Mattox:1996apj}.}, such that $\mathrm{\sqrt{TS} \geqslant 2 \sigma}$, and a minimum number of \gray{} photons above background of $\mathrm{\gamma \geqslant 2}$; otherwise a $\mathrm{95\%}$ confidence--level upper limit is calculated. For each SED we also calculate and show the $\mathrm{1\sigma}$ uncertainty band. For each radio galaxy we initially only considered the spectral model given in the 4FGL catalogue as a description of the high-energy \gray{} emission. In the NGC\,1218 SED (Figure \ref{fig:seds-setB}) we see some tension between a power-law model description and the highest energy bin upper limit. As a result we also considered a log-parabola model for NGC\,1218, and find that with a test statistic value of 235 between the log-parabola and power-law models, the fit significantly improves and hence we discard the initial power-law model. 

\begin{figure*}
\centering
\captionbox[Text]{Spectral energy distributions obtained for the radio galaxies PKS\,1304-215, PKS\,1514+00, PKS\,1839-48, PKS\,2153-69, PKS\,2300-18, PKS\,2324-2, PKS\,2338-295, Pictor\,A, TXS\,1303+114 and TXS\,1516+064. Apart from PKS\,2153-69, all the radio galaxy SEDs in this subset are best-fitted with a simple power-law model. The binning scheme is 2 bins per decade and a $\mathrm{95\%}$ confidence--level upper limit is shown for bins where $\mathrm{\sqrt{TS} < 2 \sigma}$ and the number of \gray{} photons above background in each bin is $\mathrm{\gamma < 2}$. \label{fig:seds-setC}}{%
\includegraphics[width=0.9\textwidth]{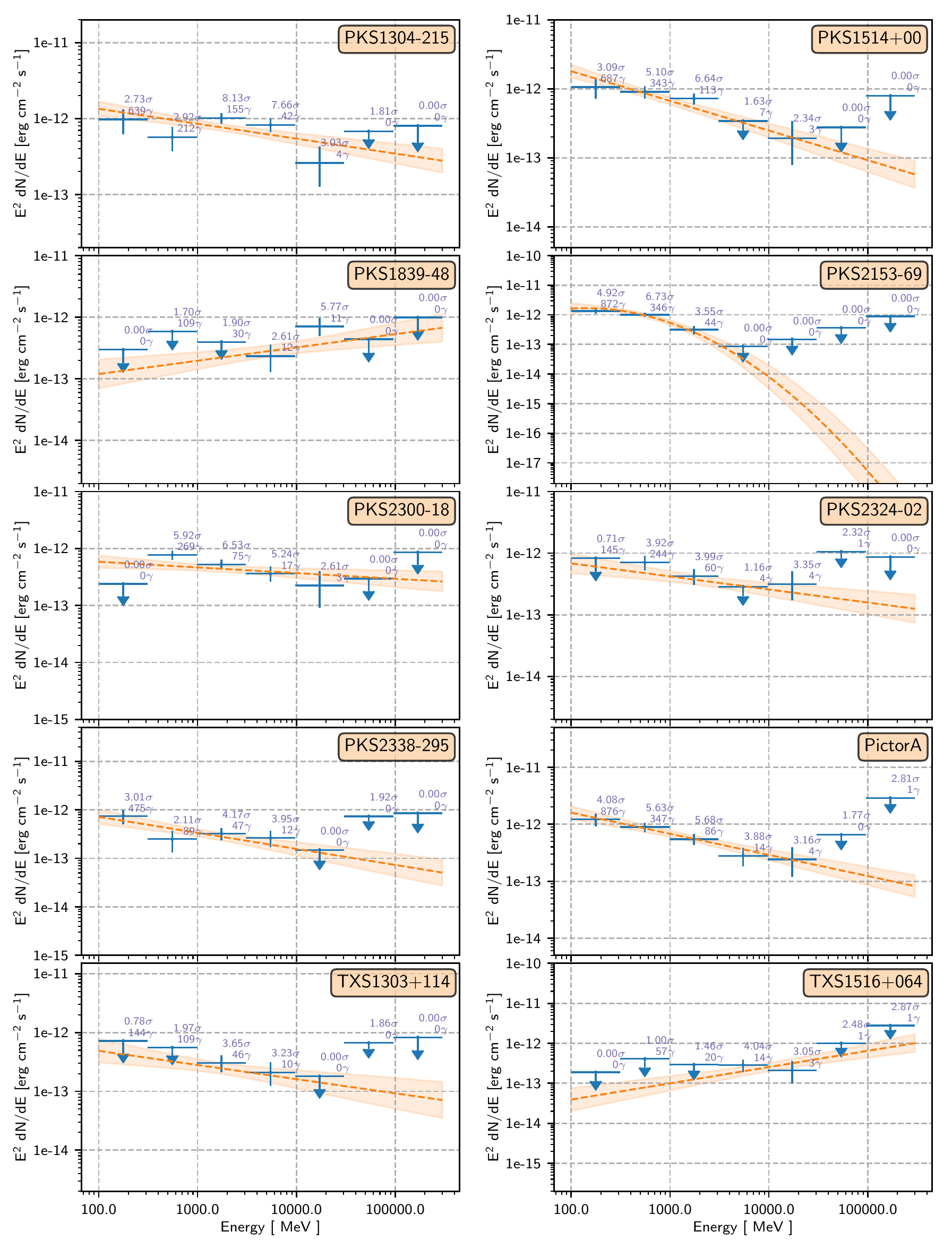}
}
\end{figure*}

\begin{table*}
\footnotesize
\begin{tabular}{| l | l | r | r | r | r |}
\hline
\fermi{} name & Assoc. name & $\mathrm{N_{0}}$ ($\mathrm{cm^{-2} \enskip s^{-1} \enskip MeV^{-1}}$) & Index ($\Gamma$) & $\mathrm{E_{0}}$ (MeV) & $\mathrm{E_{pivot}}$ (MeV) \\ \hline
4FGL J0418.2+3807 & 3C 111 & $\text{(}7.39 \pm 0.46\text{)} \times 10^{-12}$ & $-2.75 \pm 0.05$ & 532.8 & 300.0 \\ \hline
4FGL J1144.9+1937 & 3C 264 & $\text{(}2.84 \pm 0.39\text{)} \times 10^{-14}$ & $-1.90 \pm 0.10$ & 3216.4 & 3601.2 \\ \hline
4FGL J1443.1+5201 & 3C 303 & $\text{(}1.22 \pm 0.24\text{)} \times 10^{-14}$ & $-1.94 \pm 0.14$ & 3426.4 & 3250.7 \\ \hline
4FGL J0334.3+3920 & 4C +39.12 & $\text{(}7.90 \pm 1.48\text{)} \times 10^{-15}$ & $-1.75 \pm 0.13$ & 5679.6 & 6777.7 \\ \hline
4FGL J1449.5+2746 & B2 1447+27 & $\text{(}7.23 \pm 2.49\text{)} \times 10^{-16}$ & $-1.40 \pm 0.20$ & 11380.8 & 13137.0 \\ \hline
4FGL J0322.6-3712e & Fornax A & $\text{(}2.02 \pm 0.15\text{)} \times 10^{-13}$ & $-2.11 \pm 0.06$ & 1762.5 & 1396.5 \\ \hline
4FGL J0009.7-3217 & IC 1531 & $\text{(}6.88 \pm 1.22\text{)} \times 10^{-14}$ & $-2.2 \pm 0.13$ & 1692.2 & 1440.7 \\ \hline
4FGL J1454.1+1622 & IC 4516 & $\text{(}6.76 \pm 0.63\text{)} \times 10^{-13}$ & $-2.31 \pm 0.07$  & 922.3 & 1073.6 \\ \hline
%%4FGL J0308.4+0407 & NGC 1218 & $\text{(}1.32 \pm 0.11\text{)} \times 10^{-13}$ & $-1.94 \pm 0.06$ & 2411.4 & 2628.0 \\ \hline
4FGL J0708.9+4839 & NGC 2329 & $\text{(}6.32 \pm 1.52\text{)} \times 10^{-15}$ & $-1.69 \pm 0.15$ & 4693.8 & 7416.6 \\ \hline
4FGL J0758.7+3746 & NGC 2484 & $\text{(}9.88 \pm 2.82\text{)} \times 10^{-15}$ & $-1.96 \pm 0.17$  & 3421.0 & 3346.8 \\ \hline
4FGL J0931.9+6737 & NGC 2892 & $\text{(}2.40 \pm 0.19\text{)} \times 10^{-13}$ & $-2.23 \pm 0.06$ & 1459.5 & 1205.1 \\ \hline
4FGL J0057.7+3023 & NGC 315 & $\text{(}2.56 \pm 0.35\text{)} \times 10^{-13}$ & $-2.31 \pm 0.11$ & 1124.0 & 1158.9 \\ \hline
4FGL J0627.0-3529 & PKS 0625-35 & $\text{(}2.32 \pm 0.11\text{)} \times 10^{-13}$ & $-1.88 \pm 0.04$ & 2337.5 & 2215.6 \\ \hline
4FGL J1306.7-2148 & PKS 1304-215 & $\text{(}9.91 \pm 1.16\text{)} \times 10^{-14}$ & $-2.20 \pm 0.08$ & 2151.2 & 1640.5 \\ \hline
4FGL J1516.5+0015 & PKS 1514+00 & $\text{(}8.55 \pm 1.13\text{)} \times 10^{-13}$ & $-2.43 \pm 0.10$ & 744.0 & 852.7 \\ \hline
4FGL J1843.4-4835 & PKS 1839-48 & $\text{(}1.09 \pm 0.27\text{)} \times 10^{-14}$ & $-1.78 \pm 0.16$ & 3893.3 & 6115.0 \\ \hline
4FGL J2302.8-1841 & PKS 2300-18 & $\text{(}8.58 \pm 1.22\text{)} \times 10^{-14}$ & $-2.10 \pm 0.09$ & 1788.0 & 1530.9 \\ \hline
4FGL J2341.8-2917 & PKS 2338-295 & $\text{(}7.56 \pm 1.47\text{)} \times 10^{-14}$ & $-2.33 \pm 0.15$ & 1548.7 & 1157.3 \\ \hline
4FGL J0519.6-4544 & Pictor A & $\text{(}1.72 \pm 0.23\text{)} \times 10^{-13}$ & $-2.37 \pm 0.10$ & 1462.0 & 1397.7 \\ \hline
4FGL J2326.9-0201 & PKS 2324-02 & $\text{(}1.70 \pm 0.36\text{)} \times 10^{-13}$ & $-2.21 \pm 0.13$ & 1215.0 & 2048.4 \\ \hline
4FGL J1306.3+1113 & TXS 1303+114 & $\text{(}7.06 \pm 2.16\text{)} \times 10^{-15}$ & $-2.24 \pm 0.19$ & 4187.4 & 1604.0 \\ \hline
4FGL J1518.6+0614 & TXS 1516+064 & $\text{(}3.75 \pm 1.11\text{)} \times 10^{-15}$ & $-1.59 \pm 0.19$ & 5831.5 & 10929.7 \\ \hline
\end{tabular}
\caption{Details of the power-law models best fitted to the radio galaxy SEDs.}.
\label{table:two}
\end{table*}

\begin{table*}
\footnotesize
\begin{tabular}{| l | l | r | r | r | r | r | r |}
\hline
\fermi{} name & Assoc. name & $\mathrm{N_{0}}$ ($\mathrm{cm^{-2} \enskip s^{-1} \enskip MeV^{-1}}$) & Index ($\alpha$) & Curvature ($\beta$) & $\mathrm{E_{0}}$ (MeV) & $\mathrm{E_{pivot}}$ (MeV) \\ \hline
4FGL J0433.0+0522 & 3C 120 & $ \text{(}14.8 \pm 0.86 \text{)} \times 10^{-12}$ & $2.55 \pm 0.07$ & $0.23 \pm 0.05$ & 445.8 & 366.8 \\ \hline
4FGL J0308.4+0407 & NGC 1218 & $ \text{(}7.27 \pm 0.78 \text{)} \times 10^{-13}$ & $1.76 \pm 0.10$ & $0.07 \pm 0.03$ & 1000.0 & 2518.8 \\ \hline
4FGL J1630.6+8234 & NGC 6251 & $ \text{(}4.28 \pm 0.17 \text{)} \times 10^{-12}$ & $2.26 \pm 0.04$ & $0.08 \pm 0.02$ & 667.8 & 479.9 \\ \hline
4FGL J2156.0-6942 & PKS2153-69 & $ \text{(}5.68 \pm 0.91 \text{)} \times 10^{-12}$ & $2.57 \pm 0.23$ & $0.29 \pm 0.18$ & 368.8 & 295.5 \\ \hline
\end{tabular}
\caption{Details of the log-parabola model best fitted to the radio galaxy SEDs.}.
\label{table:three}
\end{table*}

%%--------------------------------------------------
%% Do we find any CenA-like spectral features?
%%--------------------------------------------------

\begin{figure}
\centering
\includegraphics[width=0.5\textwidth]{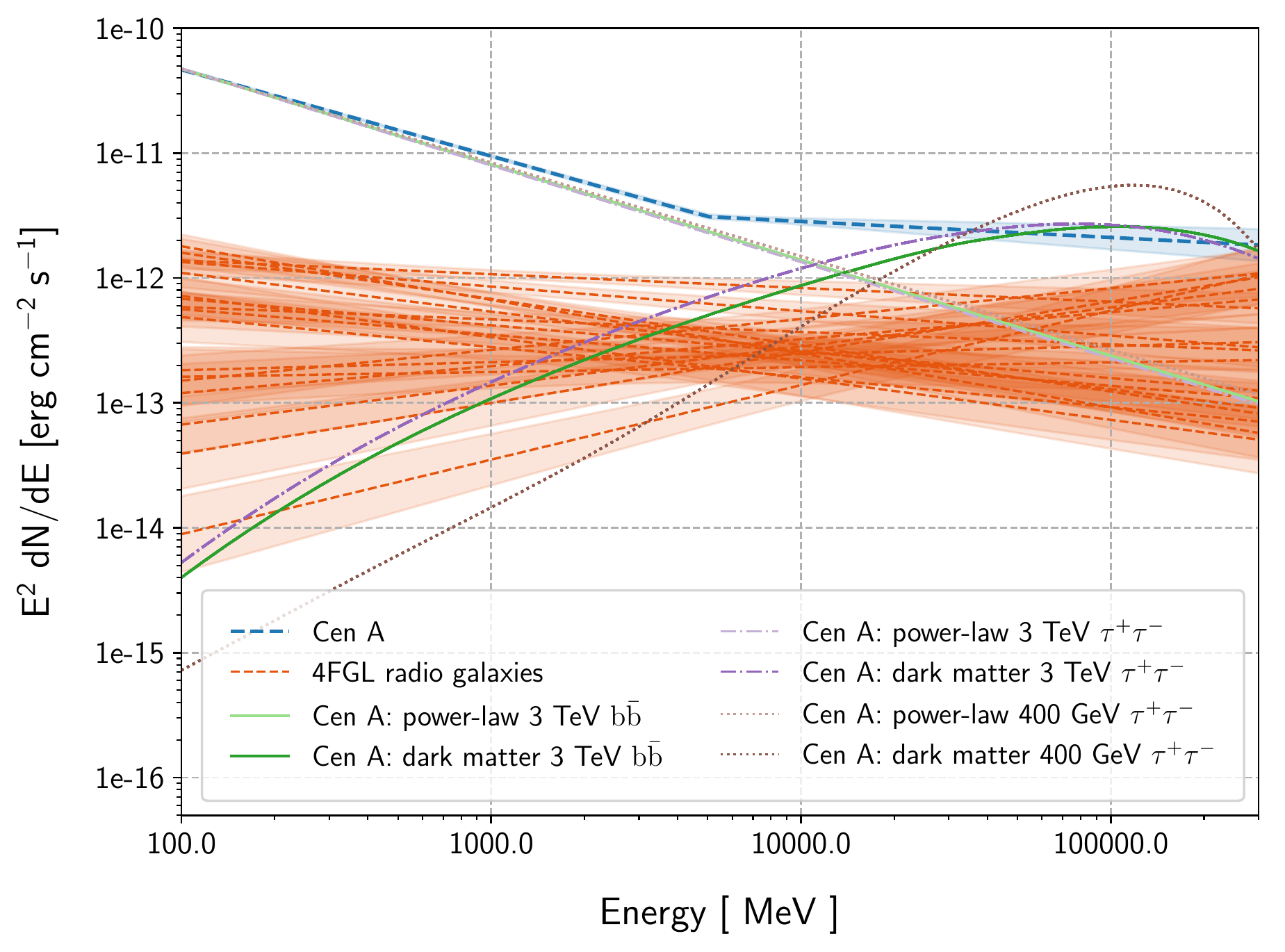}
\caption{Shown here is a comparison of the SED shapes obtained for \CenA{} (blue dashed line) versus the best fitted power-law spectral shapes for the non-variable 4FGL radio galaxies analysed in this work (orange dashed lines). Also shown is the 1 sigma confidence bands obtained for \CenA{} (blue band) and the 4FGL radio galaxies (orange bands). It is clear that if there were any \CenA{}-like spectral features present in the sample of radio galaxies analysed, we would have seen them. We have also overlaid the power-law and DM models (see legend) used to describe the total \CenA{} emission \citep{Brown-2017-PhRvD} for the given DM particle mass and annihilation channel. Assuming the supermassive black holes of the 4FGL radio galaxies are of a similar mass scale to that of \CenA's, which implies that the total mass of DM in the 4FGL radio galaxies is similar to that of \CenA{}, we see no evidence of spectral hardening in the \fermi{} energy band to warrant consideration of DM scenarios.}
\label{fig:sed-cenA-3C264-4C3912}
\end{figure}

As can be seen in Figures \ref{fig:seds-setA}, \ref{fig:seds-setB}, \ref{fig:seds-setC} and \ref{fig:sed-cenA-3C264-4C3912}, the 10-year SEDs produced do not share the spectral features which were seen in \CenA{} \citep{Brown-2017-PhRvD}. Instead, the majority of radio galaxy SEDs are best fitted with a simple power-law model. The exceptions are 3C\,120, NGC\,1218, NGC\,6251 and PKS\,2152-69, which are best fitted with a log-parabola model. We also note that NGC\,2484, PKS\,1839-48, TXS\,1303+114 and TXS\,1516+064 have only 2 statistically significant spectral flux bins above background; unsurprisingly, these targets are amongst those with the lowest detection significances within this analysis. Both NGC 2484 and TXS\,1303+114 fall just below the accepted $\mathrm{5 \sigma}$ significance threshold in this 10-year dataset, and in both cases, their SEDs lack sufficient statistics across the full energy band to produce a reliable power-law fit.

%%-------------------------------------------------------------------------
%% Do we see any other common/interesting characteristics in the RG SEDs?
%%-------------------------------------------------------------------------
In most cases we detect no significant \gray{} excess above $\mathrm{30~GeV}$. The exceptions to this are 3C\,264, 4C\,+39.12, B2\,1447+27, Fornax\,A, NGC\,1218, NGC\,2329 and PKS\,0625-35. Two of these are detected at TeV energies: PKS\,0625-35 \citep{Abdalla-2018-MNRAS} and 3C\,264 \citep{Mukherjee-2018-ATel-11436}. Apart from Fornax\,A, these radio galaxies all exhibit fairly hard spectra over the energy band considered in this analysis. In addition, 3C\,264 and 4C\,+39.21 share very similar SED characteristics across the \fermi{} energy band. Intriguingly, no significant \gray{} excess above background is detected for either of these two radio galaxies at energies below $\mathrm{1~GeV}$. Above $\mathrm{1~GeV}$ their flux brightness is similar and their spectral indices are hard (0.1 and 0.14 respectively in $\mathrm{E^{2}dN/dE}$ units), resulting in a significant \gray{} excess up to energies of $\mathrm{100~GeV}$. Given 3C 264 was recently detected by VERITAS at TeV energies \citep{Mukherjee-2018-ATel-11436}, perhaps there is potential for detecting 4C\,+39.12 at TeV energies too. If detected, it would be the lowest luminosity radio galaxy yet seen at TeV energies apart from IC\,310, a peculiar galaxy with somewhat uncertain classification \citep{Graham:2019mnras}.

In addition to this spectral investigation, we also investigated the temporal characteristics by producing lightcurves for each radio galaxy studied. Relative to blazars, radio galaxies are weak \gray{} emitting sources and at the energies we are considering, \fermi{} does not have the sensitivity performance to detect enough \gray{} photons for a well-sampled lightcurve at timescales under 6 months. Therefore we constructed lightcurves using 20 time bins with each bin comprising 180 days of \fermi{} data covering the full 10 year observation period.

The lightcurves were produced using a binned likelihood approach where the normalisation value for the radio galaxy of interest, all sources within 1 degree of the radio galaxy and the diffuse background were kept free to vary. For each time bin a flux is calculated over the full energy range $\mathrm{100\; MeV \leqslant E \leqslant 300\; GeV}$. A variability index (see Table \ref{table:one}) was calculated for each radio galaxy using the method described in \cite{Nolan-2012-ApJS} which is a simple Likelihood Ratio test between the null hypothesis (a constant source) and the alternative hypothesis (variable source). If the null hypothesis is correct, then in accordance with this method, the variability index is distributed as $\mathrm{\chi^{2}}$ with 19 degrees of freedom. Thus any variability index above 39.7 indicates evidence for variability at the $\mathrm{\geqslant 3 \sigma}$ level. We find statistical evidence of flux variability on 6 month timescales for seven of the 26 radio galaxies analysed: 3C\,111, 3C\,120, 3C\,264, IC\,4516, NGC\,1218, NGC\,2892 and PKS\,0625-35.

While fixed-time-width binning schemes, like that used in our temporal analysis are widely used within the field, there is merit in exploring a Bayesian Block binning scheme for the radio galaxies we identify here as variable. Such an analysis will aid the comparison of spectra obtained for flaring and non-flaring states and may enable us to uncover any \CenA{}-like spectral components potentially camouflaged by variable components. Such an analysis will be the subject of a future publication.

%++++++++++++++++++++++++++++++++++++++++++++++++
%
%     SECTION 4: INTERPRETATION
%
%++++++++++++++++++++++++++++++++++++++++++++++++

\section{Interpretation}
\label{sec:interpretation}
%%---------------------------------------------------
%% Why do we not see any CenA-like spectral features?
%%---------------------------------------------------
Our analysis shows that the SEDs of the radio galaxies studied do not exhibit any spectral features to warrant fitting any extra spectral components. Although no firm  conclusions can be drawn without a full multi-wavelength analysis on a source by source basis, it does appear as though the \gray{} emission found in the \fermi{} categorised radio galaxies at energies  $\mathrm{10~ MeV \leqslant E \leqslant 300~GeV}$ is dominated by jet particle acceleration and/or jet interaction. This is particularly the case for the 7 objects which display evidence for variability; such variability not only suggests a jet origin for the emission, but renders it difficult to detect any steady spectral features which may exist \citep{Graham:2019mnras}. For example, in blazars we know that the variable emission we detect is dominated by emission processes in their ultra-relativistic jets. TeV emission from blazars due to VLBI knots is seen in jets with variability timescales typically lasting minutes to hours during flaring periods. In the case of two radio galaxies, M\,87 and IC\,310, the variability timescales detected are much shorter than the light crossing time of the black hole horizon, which implies the \gray{} emission must be coming from a compact region \citep{Giannios:2009mnras}.  

%%-----------------------------------------------------------------------------------------
%% If there were any spectral features similar to those in Cen A, would we have seen them?
%%-----------------------------------------------------------------------------------------
It is immediately obvious from Figure \ref{fig:sed-cenA-3C264-4C3912} that if there were any spectral features similar to that seen in \CenA{}, they would have been detected. However, as the supermassive black hole masses for the radio galaxies are not known and the mass of the DM spike expected around the central SMBH is related to the mass of the SMBH, detailed modelling of any DM signal one might expect from these radio galaxies and how that compares to the \CenA{} DM model fitted by \cite{Brown-2017-PhRvD} is not possible; we can simply say that there is no such component at a similar level to that observed in \CenA{}. The observational evidence from \fermi{} and H.E.S.S. suggests that whatever is happening in \CenA{} is unusual, or perhaps spectral features are simply easier to detect due to the object's proximity. The lack of evidence for variability in the \gray{} emission from \CenA{} means theorists are able to postulate scenarios where the \gray{} emission arises from larger scales i.e. not a compact emitting region. Such scenarios could include contributions from undetected millisecond pulsars or dark matter \citep{Brown-2017-PhRvD} or hadronic processes such as the interaction of energetic protons with ambient matter (proton-proton interactions) \citep{Sahakyan-2013-ApJ} or the inverse Compton upscattering of photons on kiloparsec scales \citep{Hardcastle-2011-MNRAS} or host galaxy starlight \citep{Stawarz-2003-ApJ}. These scenarios are largely degenerate and the only way to distinguish these models from one another is to accumulate more and better quality radio and ground-based TeV observations of radio galaxies.\\
%
%%All we know is that we expect a radio galaxy BH mass to be of order 10^8 solar mass.
%-------------------------------------------------------------------------------------------
% Commenting on the hard spectra
%-------------------------------------------------------------------------------------------
Six of the radio galaxies studied, 3C\,264, 4C\,+39.12, B2\,1447+27, NGC\,1218, NGC\,2329 and PKS\,0625-35, have particularly hard spectra and emission above 30\,GeV. Furthermore, two of these objects, 3C\,264 and 4C\,+39.21, show no significant excess below 1\,GeV. It is possible that the hard spectra displayed by these objects are an indication that the peak of the inverse Compton emission is located in the \fermi{} energy regime. This would be surprising, as radio galaxies do not have the strong Doppler-boosting normally required to produce such a high-frequency peak in their SED. Multi-wavelength observations would be required to confirm if this is the case. 

%%-------------------------------------------------------------------------
%% If there are any other common/interesting RG characteristics, what
%% scientific interpretations can we make from these?
%%-------------------------------------------------------------------------

\begin{figure*}
\centering
\captionbox[Text]{The left panel shows the radio luminosity calculated using the total 5 GHz (where not available we used 4.8 GHz) radio flux density available in publicly accessible radio catalogues versus the \gray{} luminosity calculated using the integrated fluxes estimated in this work. We find a strong positive correlation between the radio and \gray{} luminosity. The right panel shows the absolute magnitude for the visual optical band primarily using extinction corrected V filter data (where not available we used B filter data) versus the \gray{} luminosity calculated using the integrated fluxes estimated in this work. We calculated the absolute magnitudes using photometric data available in NED and Simbad \citep{SIMBAD-2000-AAS}, except for PKS\,0625-35 \citep{Massardi-2008-MNRAS}. We find a weak correlation between the optical brightness of these radio galaxies versus their \gray{} luminosity. The TeV-detected radio galaxies are annotated and highlighted using blue star markers, and we see no clustering of these particular sources. The \CenA{} \gray{} luminosity was calculated assuming a power-law model and not the broken power-law model from \cite{Brown-2017-PhRvD}. \label{fig:correlation-analysis-one}}{%\subcaption*{Source: www...}
\includegraphics[width=0.5\textwidth]{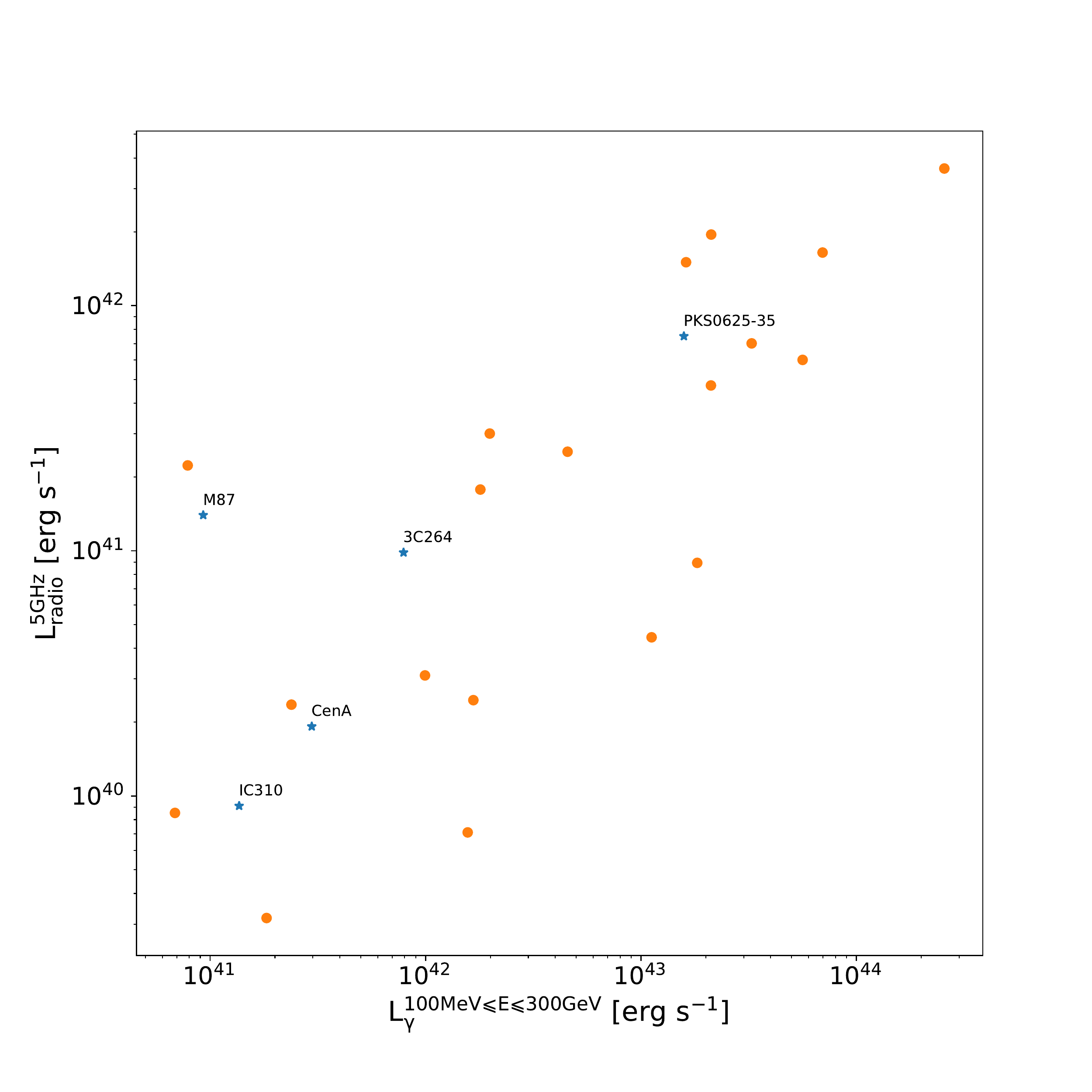}
\includegraphics[width=0.5\textwidth]{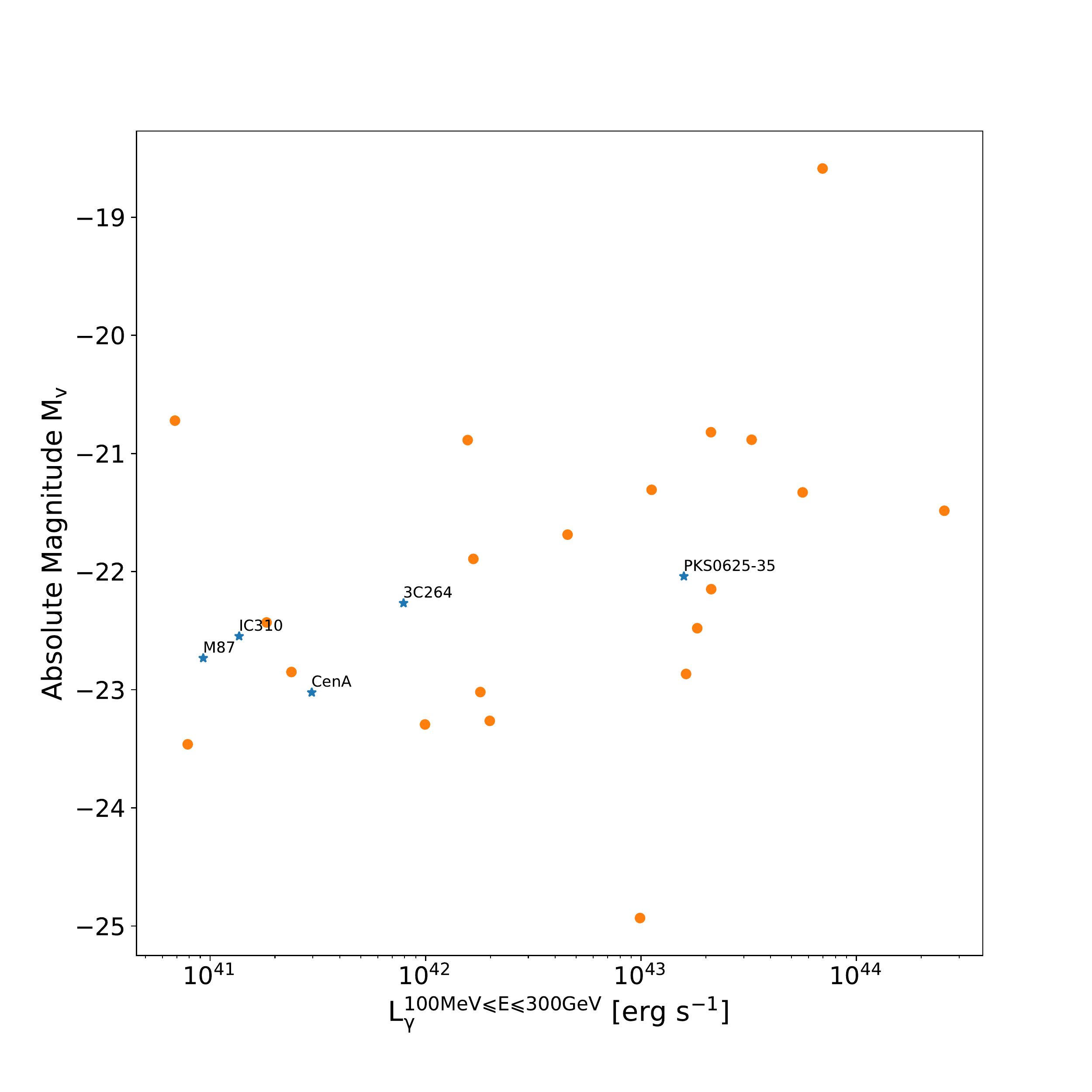}}
\end{figure*}

\begin{figure*}
\centering
\captionbox[Text]{The left panel shows the radio core power calculated using the core 5 GHz (except for PKS\,2324-02 we used 4.8 GHz) radio flux density versus the \gray{} luminosity calculated using the integrated fluxes estimated in this work. The 5 GHz radio flux densities were taken from the NASA/IPAC Extragalactic Database (NED) except for those objects with references listed below$\dagger$. We find a strong positive correlation between radio core power at 5GHz and \gray{} luminosity. The right panel shows the radio core dominance parameter versus the \gray{} luminosity calculated using the integrated fluxes estimated in this work. We find no correlation between the radio core dominance parameter of these radio galaxies versus their \gray{} luminosity. The TeV-detected radio galaxies are annotated and highlighted using blue star markers, and we see no clustering of these particular sources. $\dagger$Fornax\,A and IC\,1531 \citep{Ekers-1989-MNRAS}, NGC\,1218 \citep{Saikia-NGC1218-1986-MNRAS}, NGC\,2892 \citep{Kharb-2004-AA} and NGC\,6251 \citep{Evans-NGC6251-2005-MNRAS}. \label{fig:correlation-analysis-two}}{%\subcaption*{Source: www...}
\includegraphics[width=0.5\textwidth]{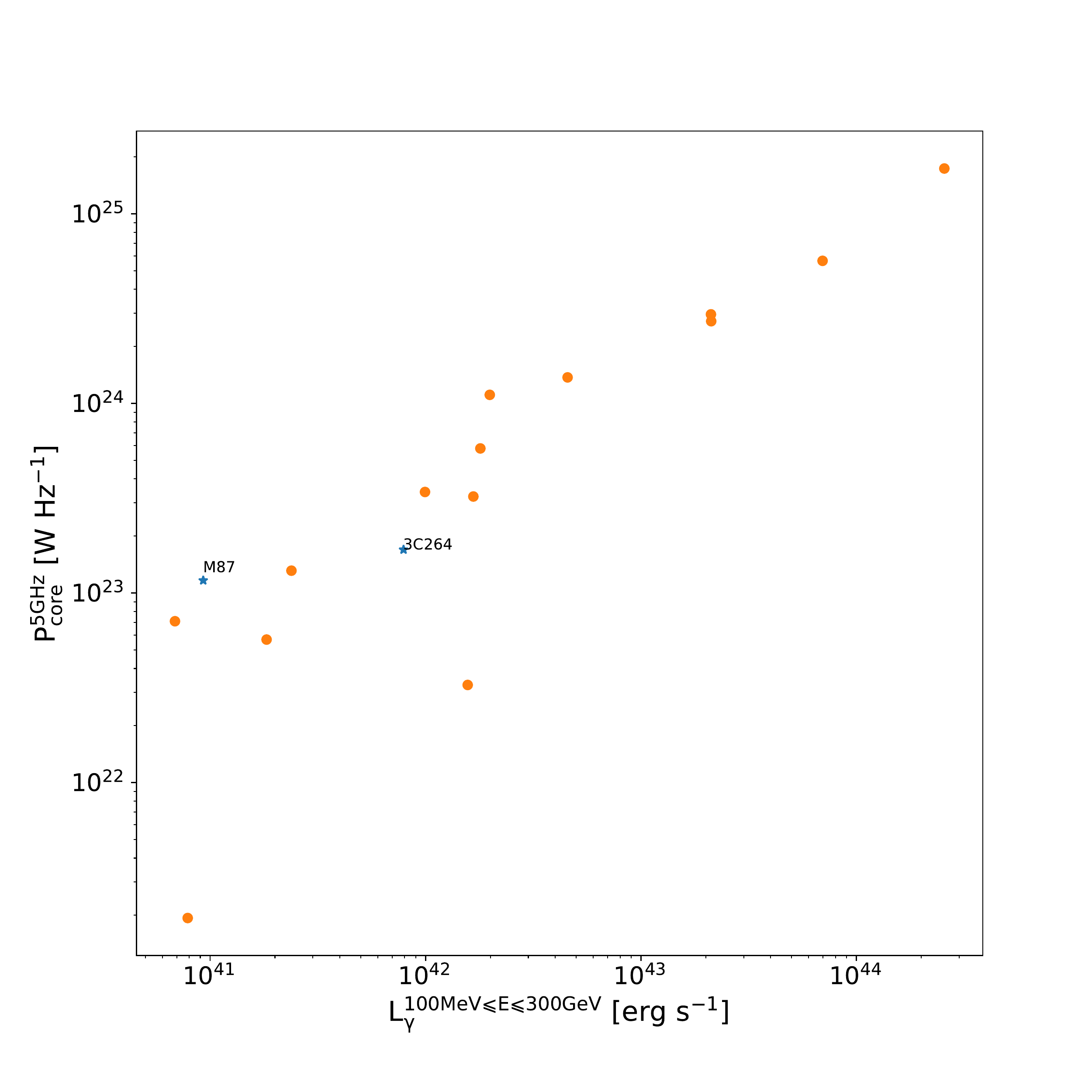}
\includegraphics[width=0.5\textwidth]{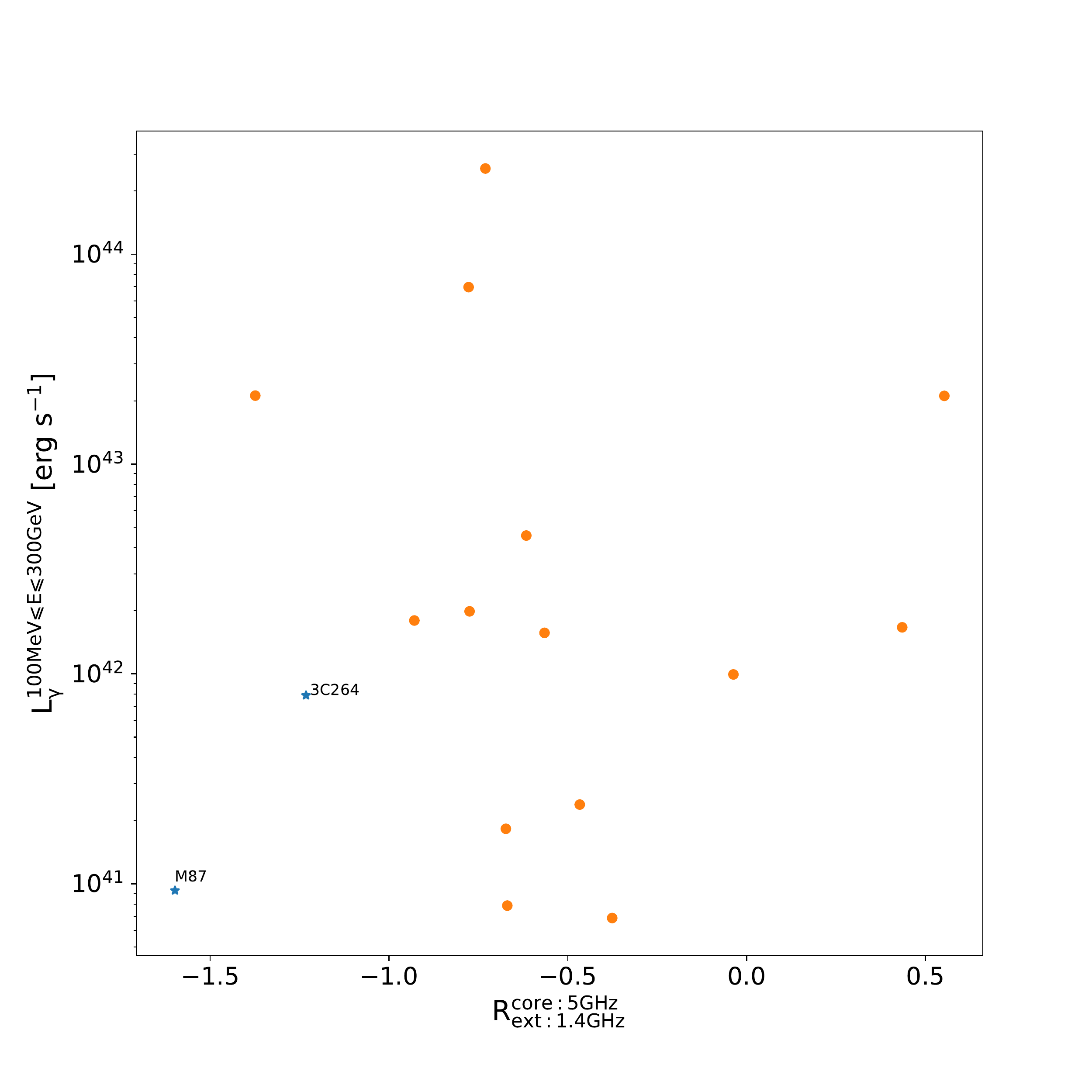}}
\end{figure*}

\begin{figure*}
\centering
\captionbox[Text]{Skymaps showing the significance $\mathrm{\sqrt{TS}}$ for a subset of the radio galaxies analysed. Each skymap considers all energies between $\mathrm{100 \enskip MeV \leq E \leq 300 \enskip GeV}$ and the intensity scale in the z-axis highlights the significance. The dark-orange solid contour line indicates the $\mathrm{5\sigma}$ significance boundary and the light-orange solid contour the $\mathrm{15\sigma}$ significance boundary. The radio galaxy position is indicated with a green $\mathrm{\times}$ and the two orange dashed-line concentric circles in the upper left corner of the top left panel show the approximate \fermi{} PSF at $\mathrm{100 \enskip MeV \enskip \textrm{(large)} \enskip and \enskip 1 \enskip GeV \enskip \textrm{(small)}}$ respectively. \label{fig:tsmaps-setA}}{%\subcaption*{Source: www...}
\includegraphics[width=0.9\textwidth]{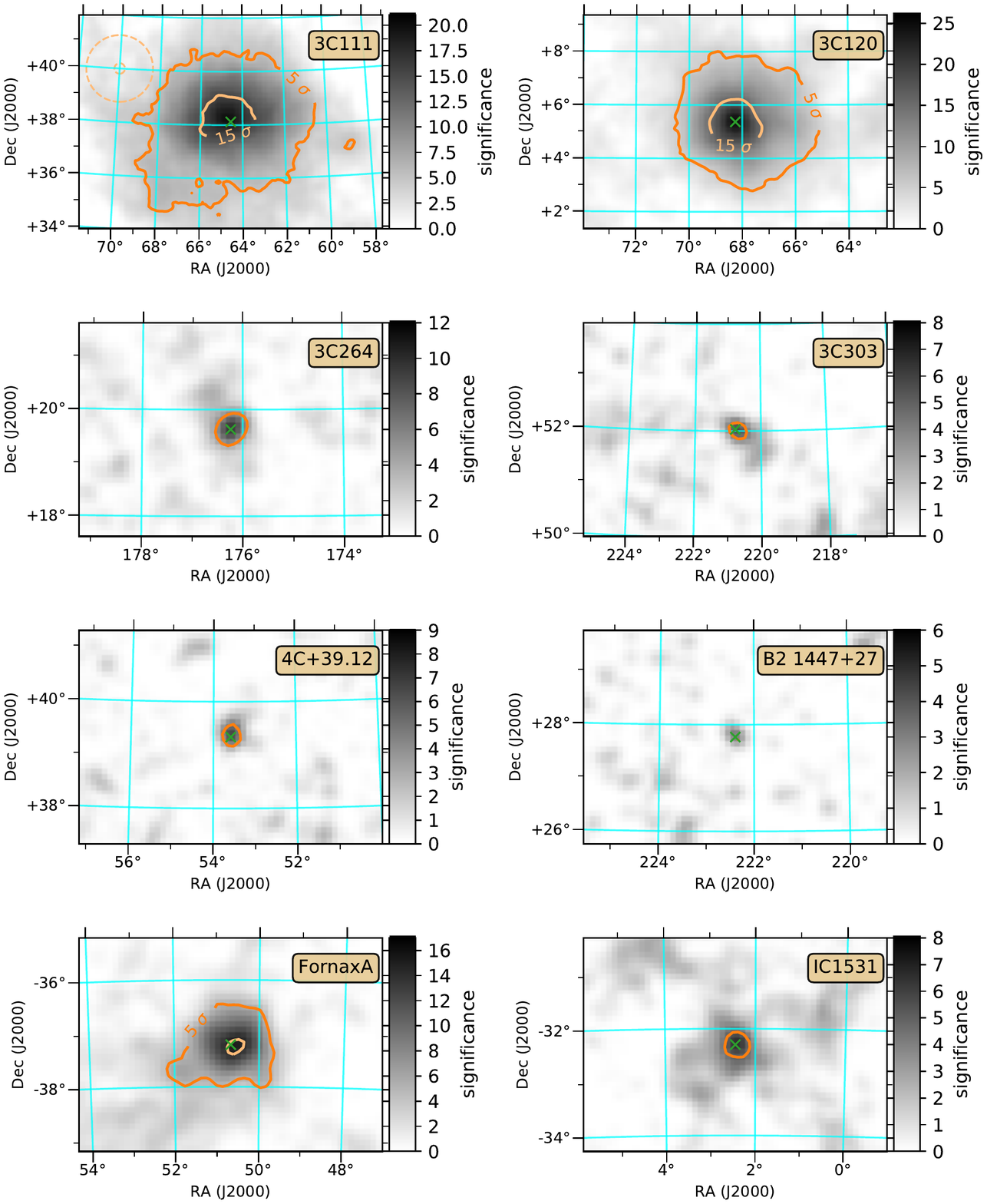}
}
\end{figure*}

\begin{figure*}
\centering
\captionbox[Text]{Skymaps showing the significance $\mathrm{\sqrt{TS}}$ for a subset of the radio galaxies analysed. Each skymap considers all energies between $\mathrm{100 \enskip MeV \leq E \leq 300 \enskip GeV}$ and the intensity scale in the z-axis highlights the significance. The dark-orange solid contour line indicates the $\mathrm{5\sigma}$ significance boundary and the light-orange solid contour the $\mathrm{15\sigma}$ significance boundary. The radio galaxy position is indicated with a green $\mathrm{\times}$ and the two orange dashed-line concentric circles in the upper left corner of the top left panel show the approximate \fermi{} PSF at $\mathrm{100 \enskip MeV \enskip \textrm{(large)} \enskip and \enskip 1 \enskip GeV \enskip \textrm{(small)}}$ respectively. \label{fig:tsmaps-setB}}{%\subcaption*{Source: www...}
\includegraphics[width=0.9\textwidth]{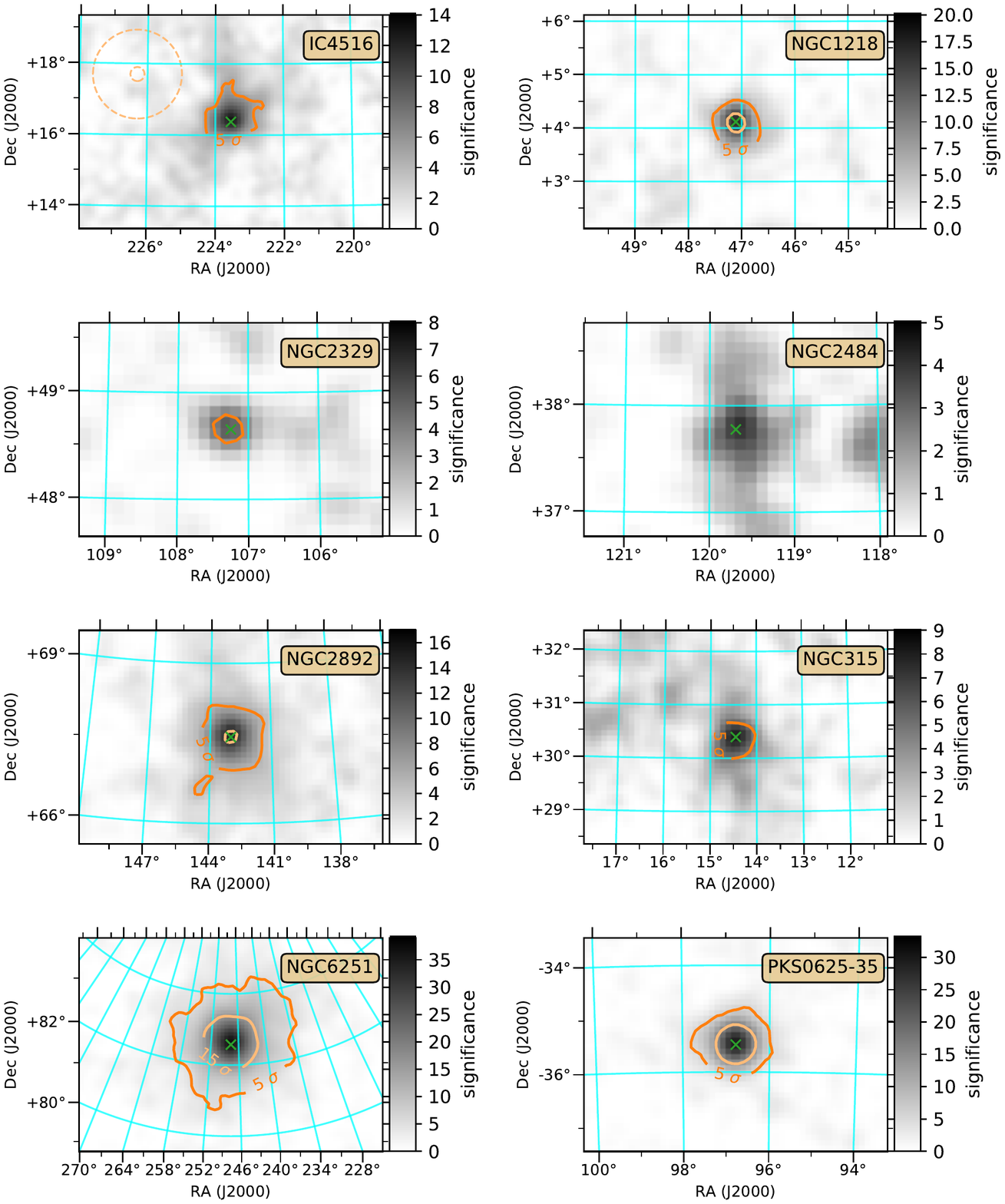}
}
\end{figure*}

\begin{figure*}
\centering
\captionbox[Text]{Skymaps showing the significance $\mathrm{\sqrt{TS}}$ for a subset of the radio galaxies analysed. Each skymap considers all energies between $\mathrm{100 \enskip MeV \leq E \leq 300 \enskip GeV}$ and the intensity scale in the z-axis highlights the significance. The dark-orange solid contour line indicates the $\mathrm{5\sigma}$ significance boundary and the light-orange solid contour the $\mathrm{15\sigma}$ significance boundary. The radio galaxy position is indicated with a green $\mathrm{\times}$ and the two orange dashed-line concentric circles in the upper left corner of the top left panel show the approximate \fermi{} PSF at $\mathrm{100 \enskip MeV \enskip \textrm{(large)} \enskip and \enskip 1 \enskip GeV \enskip \textrm{(small)}}$ respectively. \label{fig:tsmaps-setB}}{%\subcaption*{Source: www...}
\includegraphics[width=0.9\textwidth]{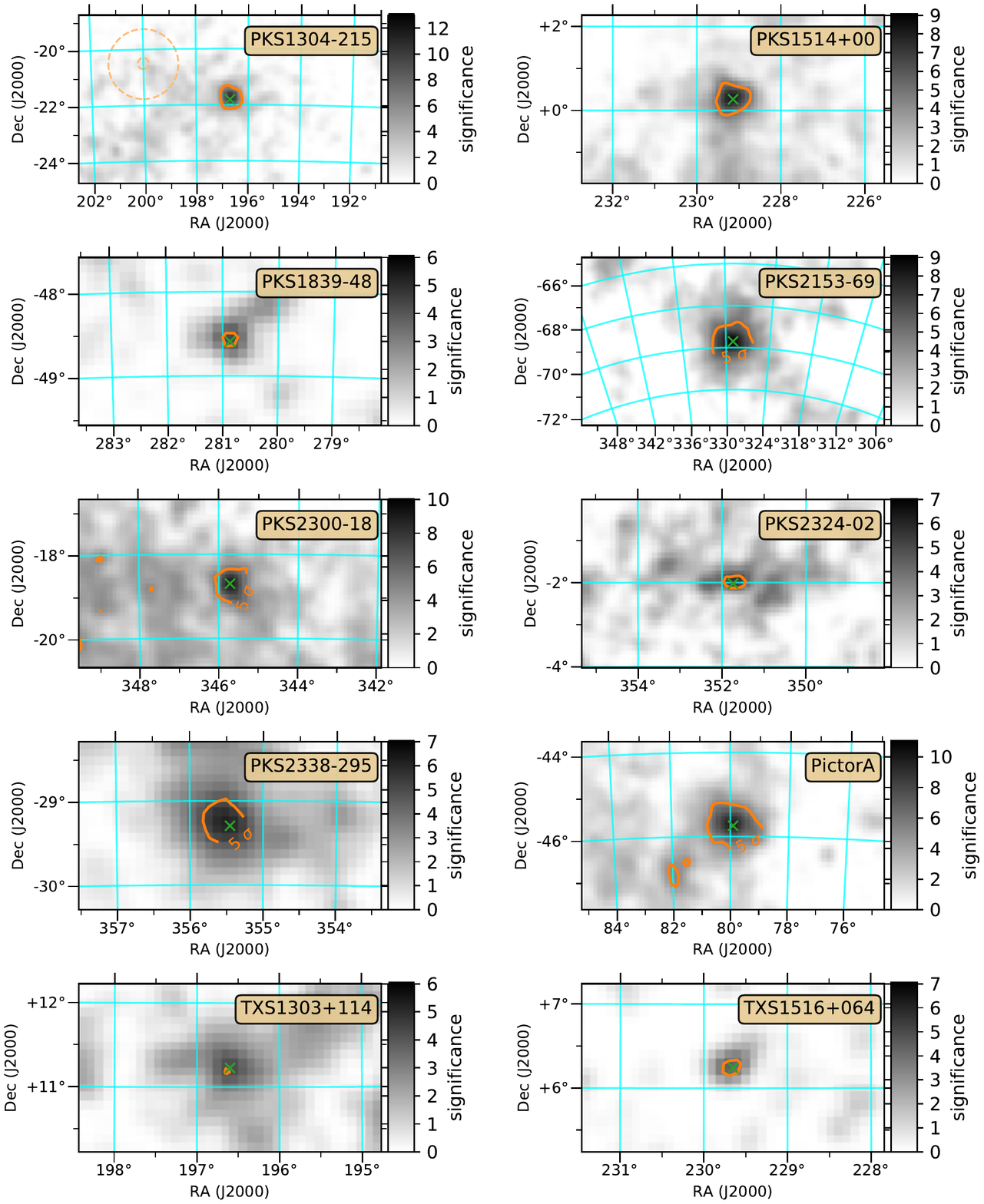}
}
\end{figure*}

%%-------------------------------------------------------------------------
%% What can we say about any correlations we find for these radio galaxies
%%-------------------------------------------------------------------------

In the case of \CenA{}, the H.E.S.S. observations were really important in the identification of a statistically significant spectral hardening. Thus, with the construction of new ground-based instruments like CTA about to begin, we need to look for any hints of correlation across multiwavelength data to try and pinpoint the best radio galaxy candidates for observation with IACTs, which ultimately will help us to better understand any such spectral upturns and if they are a common occurrence in these galaxies. We therefore used our analysis results to search for correlations between a number of different characteristic properties and the \gray{} emission from these radio galaxies. %We searched for correlations between a number of different characteristic properties and the \gray{} emission from these radio galaxies.
Figures \ref{fig:correlation-analysis-one} and \ref{fig:correlation-analysis-two} show the results of our correlation studies using data from publicly accessible radio and optical catalogues.

In Figure \ref{fig:correlation-analysis-one} the left panel shows the radio luminosity calculated using the \textit{total} 5 GHz radio flux density versus the \gray{} luminosity calculated using the integrated fluxes estimated in this work, assuming a concordance cosmology with $\mathrm{H_{0} = 71\; km\; s^{-1}\; Mpc^{-1}}$, $\mathrm{\Omega_{M} = 0.27}$, ${\mathrm{\Omega_{\Lambda}=0.73}}$ and $\mathrm{T_{CMB} = 2.725\; K}$. Where no 5 GHz flux densities were available we used the 4.8 GHz flux density measurement. We find a strong positive correlation between the radio and \gray{} luminosity, with the correlation coefficient $\mathrm{r = 0.8}$. The right panel shows the absolute magnitude of the visual optical band calculated using extinction corrected V filter data versus the \gray{} luminosity, calculated using the integrated fluxes estimated in this work. Where no V filter data were available we used B filter data. We find a weak positive correlation ($\mathrm{r = 0.4}$) between the absolute magnitude of these radio galaxies versus their \gray{} luminosity. We also highlight and annotate the power-law modelled TeV-detected radio galaxies using blue star markers, and we see no clustering of these particular sources. 

The left panel of Figure \ref{fig:correlation-analysis-two} shows the radio power of the \textit{core} at 5 GHz frequencies versus the \gray{} luminosity calculated using the integrated fluxes estimated in this work. Where no 5 GHz data were available we used the 4.8 GHz radio flux densities to estimate the core radio power. We find a strong positive correlation between the core radio power at 5 GHz frequencies versus the \gray{} luminosity, with the correlation coefficient $\mathrm{r = 0.9}$. The right panel shows the \gray{} luminosity calculated using the integrated fluxes estimated in this work versus the radio core dominance parameter calculated using the method highlighted in \cite{Fan-Zhang-2003-AA}. The core dominance parameter suffers from a number of different systematic uncertainties \citep{Abdo-Misaligned-2010-ApJ}, and as reported elsewhere \citep{Angioni-2019-AA} we find no correlation between the \gray{} luminosity and the radio core dominance parameter, with correlation coefficient $\mathrm{r = -0.3}$. Again we highlight and annotate the TeV-detected radio galaxies using blue star markers, and we see no clustering of these particular sources.

%%-----------------------------------------------------------------------------------------
%% Comment on evidence for extension - probably another paper :-)
%%-----------------------------------------------------------------------------------------

Finally, in the context of potentially correlated $\gamma$-ray and radio flux, we also investigated the possibility of extended $\gamma$-ray emission associated with the kiloparsec scale jet of the radio galaxies. For each radio galaxy we produced a skymap as seen in Figures \ref{fig:tsmaps-setA} and \ref{fig:tsmaps-setB}. Each of these show the significance ($\mathrm{\sqrt{TS}}$) for an approximate 2 degree region centred on the radio galaxy target (indicated with a green $\mathsf{x}$). Significance values greater than 5 $\sigma$ are enclosed within the solid dark-orange contour line, and values greater than 15 $\sigma$ are enclosed within the solid light-orange contour line. We note that there is apparent evidence for extended emission coming from the direction of 3C\,111. However, on closer inspection this extension is likely an artefact of nearby ($\mathrm{<1.5^{\circ}}$) point sources just below the detection threshold.

%%Still to do for interpretation section:
% 1. What do the correlations tell us?
% 2. I suspect we may just have a selection effect going on here and ideally we need to compare against all 4FGL AGN?
% 3. For DM studies suggest looking at non-core dominated RGs to look for gamma-ray emission. A plug and lead into Max's analysis/work.

%%-------------------------------------------------------------------------
%% What can we say / speculate about the lack of low energy emission
%% from 3C264 and 4C+39.12
%%-------------------------------------------------------------------------

%%-------------------------------------------------------------------------
%% What can we say about these radio galaxies and the prospects for
%% observing them with CTA
%%-------------------------------------------------------------------------

\begin{figure*}
\centering
{%\subcaption*{Source: www...}
\includegraphics[width=0.9\textwidth]{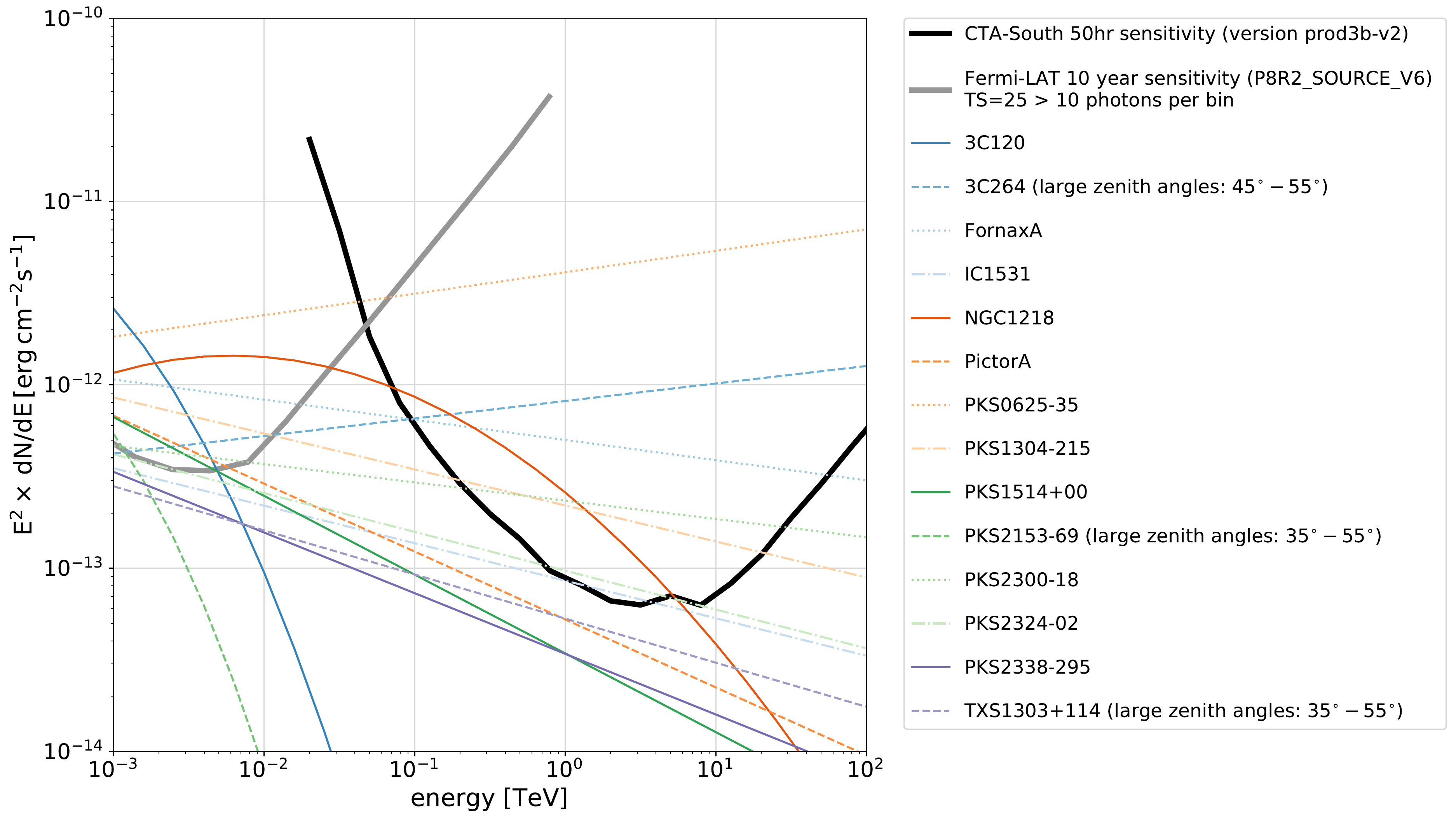}
}
\captionbox[Text]{Shown here are the \fermi{} 4FGL radio galaxy fluxes extrapolated up to 100 TeV for both CTA-South (top panel) and CTA-North (bottom panel). The respective sensitivity performance curves for each of the CTA sites is also shown (solid black line) as well as the \fermi{} 10 year sensitivity (solid grey line). The \fermi{} detected fluxes are extrapolated assuming no spectral breaks or features between the GeV and TeV energy range. \label{fig:extrapolation-south}}
{
\includegraphics[width=0.9\textwidth]{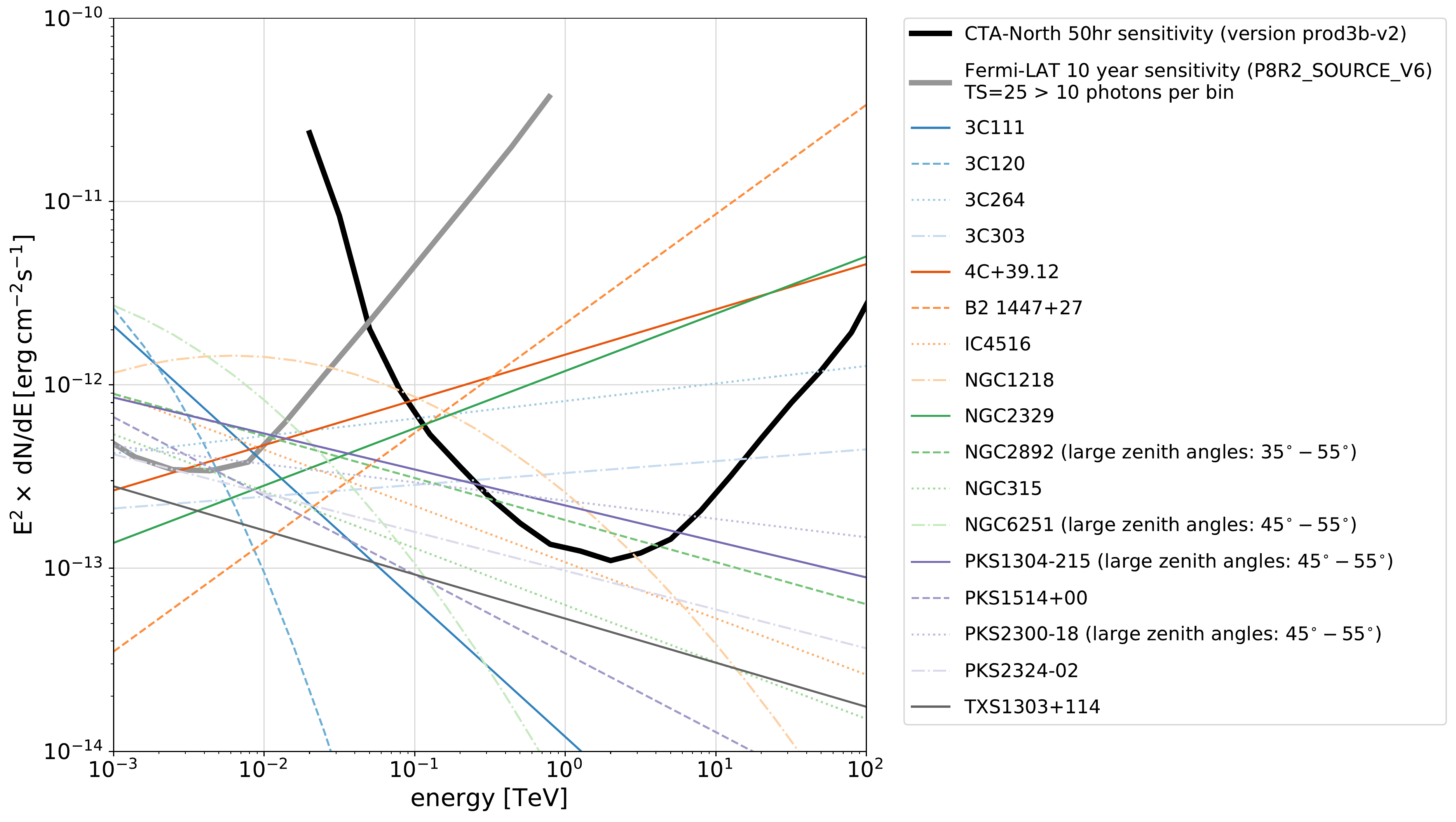}
}
\end{figure*}

As discussed above, radio galaxies are very interesting targets for a host of reasons. With the forthcoming next-generation ground-based \gray{} observatory the Cherenkov Telescope Array (CTA) (\cite{CTA-Ong-2019-EPJWC}; \cite{Angioni-CTA-2017}), it is hoped that a larger sample of radio galaxies emitting radiation at very-high-energies will be gathered. Figure \ref{fig:extrapolation-south} shows the \fermi{} 4FGL radio galaxy fluxes extrapolated to 100 TeV for both CTA-South (left panel) and CTA-North (right panel) respectively. The \fermi{} detected fluxes were extrapolated assuming no breaks or features in the spectra from the GeV to TeV energy regime. CTA's  ten times better sensitivity over the core energies compared to existing ground-based instruments should enable the detection of approximately 13 of the radio galaxies analysed in this work, assuming a 50 hour observation using the CTA-North and CTA-South arrays respectively. In practice, CTA has the potential to detect a larger number, as there are bound to be variable radio galaxies that will be seen with CTA during flaring periods. 

%++++++++++++++++++++++++++++++++++++++++++++++++
%
%     SECTION 5: CONCLUSIONS
%
%++++++++++++++++++++++++++++++++++++++++++++++++

%%-------------------------------------------------------------------------
%% What can we conclude from this work? Where do we go from here?
%%-------------------------------------------------------------------------
\section{Conclusions}
\label{sec:conclusions}

 The discovery of a distinctive break in the power-law spectrum of \CenA{} posed many questions concerning the origin and mechanisms behind the spectral hardening. We therefore analysed $\mathrm{26}$ other \fermi-detected radio galaxies to see whether any other ``similar'' objects to \CenA{} exhibit breaks and spectral hardening. This work has found no evidence for spectral hardening over a 10-year-averaged spectrum calculated for each radio galaxy. Had there been such a spectral feature in these objects, it would have been apparent in the data we analysed. This suggests that either \CenA{} is unique among radio galaxies, or that any break occurs outside \fermi's energy range. We also noted that a number of the galaxies analysed show variability on the 6-month timescale, which strongly suggests a jet origin for the \gray{} emission from these objects and would render the detection of any spectral break with a non-jet origin difficult, if not impossible. 
 
 %This leads us to conclude that the hardening seen in \CenA{} is most likely due to the acceleration of a particle population either at the base of the jet close to the core or in the jet itself. It is unlikely that the \grays{} seen from radio galaxies result from the annihilation of weakly interacting massive particles, instead acceleration of particles in the jet or shocks are the most likely causes of \grays{} via inverse Compton processes.

With the advent of new large observatories such as SKA and CTA, a new era of astronomy is upon us that can help to better understand the non-thermal astroparticle physics at play in AGN-like radio galaxies. Many open questions still remain, such as where and how \grays{} are produced in these extragalactic objects and why a fraction of these largest and most energetically connected objects seen in our universe produce TeV \grays{} despite having much lower Doppler boosting factors compared to blazars? Although we may not get a complete understanding of how these objects work, new discoveries and findings may help to further our knowledge of the characteristics that distinguish between classes of objects under unification schemes, for example multi-wavelength studies across the broad electromagnetic spectrum from low frequency radio observations through to very-high-energy \gray{} observations may provide further support that these unifying characteristics are really down to differences in the masses of their supermassive black holes and their spins, their accretion rates, and the angles and distances at which we view these fascinating objects.

%++++++++++++++++++++++++++++++++++++++++++++++++
%
%     ACKNOWLEDGEMENTS
%
%++++++++++++++++++++++++++++++++++++++++++++++++

\section*{Acknowledgements}

The authors would like to acknowledge the excellent data and analysis tools provided by the NASA \fermi{} collaboration, without which this work could not be done. This research has made use of the CTA instrument response functions provided by the CTA Consortium and Observatory; see http://www.cta-observatory.org/science/cta-performance/ (version prod3b-v2) for more details. In addition, this research has made use of the NASA/IPAC Extragalactic Database (NED) which is operated by the Jet Propulsion Laboratory, California Institute of Technology, under contract with the National Aeronautics and Space Administration. This research has also made use of the SIMBAD database, operated at CDS, Strasbourg, France.

Finally, the authors acknowledge the financial support of the UK Science and Technology Facilities Council consolidated grant ST/P000541/1.

%%%%%%%%%%%%%%%%%%%%%%%%%%%%%%%%%%%%%%%%%%%%%%%%%%

%%%%%%%%%%%%%%%%%%%% REFERENCES %%%%%%%%%%%%%%%%%%

% The best way to enter references is to use BibTeX:

\bibliographystyle{mnras}
\bibliography{RGs-references}
%\bibliography{example} % if your bibtex file is called example.bib

% Alternatively you could enter them by hand, like this:
% This method is tedious and prone to error if you have lots of references
% \begin{thebibliography}{99}
% \bibitem[\protect\citeauthoryear{Author}{2012}]{Author2012}
% Author A.~N., 2013, Journal of Improbable Astronomy, 1, 1
% \bibitem[\protect\citeauthoryear{Others}{2013}]{Others2013}
% Others S., 2012, Journal of Interesting Stuff, 17, 198
% \end{thebibliography}

%%%%%%%%%%%%%%%%%%%%%%%%%%%%%%%%%%%%%%%%%%%%%%%%%%

%%%%%%%%%%%%%%%%% APPENDICES %%%%%%%%%%%%%%%%%%%%%

% \appendix

% \section{Some extra material}

% If you want to present additional material which would interrupt the flow of the main paper,
% it can be placed in an Appendix which appears after the list of references.

%%%%%%%%%%%%%%%%%%%%%%%%%%%%%%%%%%%%%%%%%%%%%%%%%%

% Don't change these lines
\bsp	% typesetting comment
\label{lastpage}
\end{document}